\documentclass[12pt]{article}
\usepackage{amssymb,amsmath,epsfig,float,xcolor,booktabs,cite}
\allowdisplaybreaks
\begin{document}
\title{\bf Modeling Gravastar Structure admitting Kuchowicz Spacetime in Rastall Gravity Theory}
\author{Tayyab Naseer$^{1,2}$ \thanks{tayyab.naseer@math.uol.edu.pk;
tayyabnaseer48@yahoo.com}~, M. Sharif$^{1}$ \thanks{msharif.math@pu.edu.pk}~ and
Areej Tabassum$^{1}$ \thanks{ireej1999@gmail.com}\\
$^{1}$Department of Mathematics and Statistics, The University of Lahore,\\
1-KM Defence Road Lahore-54000, Pakistan.\\
$^{2}$Research Center of Astrophysics and Cosmology, Khazar University, \\
Baku, AZ1096, 41 Mehseti Street, Azerbaijan.}

\date{}
\maketitle
\begin{abstract}
This paper investigates the gravastar model as a potential
alternative to black holes, utilizing the Kuchowicz metric in the
context of Rastall gravity. The model comprises three distinct
regions: an interior with positive energy density and negative
pressure, a thin intermediate shell made of ultra-relativistic stiff
fluid, and an exterior vacuum. The negative pressure within the
interior generates an outward force exerted on the shell, fulfilling
the Zel'dovich criterion. This configuration eliminates the central
singularity and replaces the event horizon with the shell. We then
derive the radial metric functions for both the inner and thin
region, yielding a non-singular solution. Furthermore, we examine
the physical properties of the shell, such as its energy, proper
length, entropy, equation of state parameter, gravitational redshift
and adiabatic index, across a range of Rastall parameter values. We
conclude that the resulting gravastar model offers a promising
solution to the singularity problem of conventional black holes
within the context of this non-conservative theory.
\end{abstract}

\section{Introduction}

Cosmic structures, encompassing both smaller and massive celestial
objects, are essential in influencing the development of the
universe and serves as a basic for studies in the field of
astrophysics. To understand the mechanism and structure of these
stellar objects, various theories have been proposed. Einstein's
theory of general relativity (GR) examines the dynamic interactions
between space, time, matter, and curvature. Edwin Hubble's
observation that galaxies are moving away from us, leading to the
accelerated expansion of the universe, is supported by extensive
observational evidences. This acceleration is attributed to dark
energy, a mysterious force thought to exert enormous negative
pressure. Numerous models have been developed to explain the
universe's origin, evolution, and transitions through different
cosmic epochs \cite{1,1a}. The Big Bang theory, for example,
suggests that the universe began as a singularity, a point of
infinite temperature and energy density. While this theory is widely
regarded, alternative hypotheses, such as the Big Bounce theory,
have also been proposed. This theory suggests that the universe
undergoes continuous cycles of expansion and contraction, with no
definitive beginning or end.

From a cosmological perspective, GR serves as a fundamental
framework for understanding stellar structures, incorporating the
concept of the cosmological constant that governs the accelerated
expansion of the universe. However, despite the significant success
of GR, there remains a notable need for modifications due to various
issues associated with the cosmic expansion. One well-known aspect
of GR is the covariant conservation of the energy-momentum tensor
(EMT), which implies that the total mass of a system remains
conserved. Yet, this principle has not been experimentally verified.
As a result, alternative theories have been proposed that do not
strictly adhere to this conservation. In 1972, Rastall \cite{4a,5a}
introduced a modified theory of gravity, addressing these issues by
altering the conservation law in GR. Rastall's theory explores the
impact of quantum fields in curved spacetime in a covariant manner.
These foundational and cosmological aspects of Rastall's theory
motivate us to investigate analytical solutions within this
framework.

Rastall's theory offers a clear and manageable formulation for the
Einstein field equations (FEs), revealing noteworthy characteristics
from both cosmological and astrophysical viewpoints. Visser
\cite{50} proposed a modified theory with a non-conserved EMT and
non-minimal coupling, but it is ultimately equivalent to GR, with
its EMT. Contrary to Visser's claim, Darabi et al. \cite{51} argued
that Rastall theory is not equivalent to GR, as it is a more
flexible theory capable of addressing challenges in observational
cosmology and quantum gravity. Moreover, this modified theory has
been extensively reviewed by several researchers
\cite{40,41,42,43,44}. Debnath \cite{47} analyzed a charged
gravastar model with three regions and investigates the impact of
the Rastall parameter and Rainbow function on the system's stability
and characteristics. In another study, Shahzad and Abbas \cite{49}
examined anisotropic compact stars in using Krori-Barua metric,
focusing on their physical properties and compliance with the
Buchdahl limit. Abbas and Majeed \cite{49a} presented a new
gravastar model with isotropic matter distribution, with
singularity-free and horizon free solutions. Lately, Shababi et al.
\cite{49b} investigated the phase-space analysis of generalized
Rastall theory, showing that it supports a stable critical point for
the universe's late-time accelerated expansion. A large body of
literature exists in this theory that discussed its cosmological and
astrophysical implications
\cite{r1mb,r1o,r1w,r1x,r1h,r1y,r1z,r1l,r1n,r1p,r1c,r1a,r1b,r1j,r1m,r1d,r1da,r1e,r1ad,r1q,r1db,
19c,r1ma,r1ai,r1aa,r1ah,r1v}.

Galaxies are largely made up of stars, which are organized in a vast
cosmic network. These stars, predominantly composed of helium and
hydrogen, maintain internal stability by counteracting gravitational
forces through the ongoing process of nuclear fusion. However, when
a star depletes its fuel supply, the pressure that supports it from
inner side decreases, leading the star to collapse due to its
gravitational pull. This collapse can result in the creation of
highly dense and compact objects. Among the various objects, one of
the most compact is the black hole, a body that has collapsed to an
extreme degree. The structure of a black hole includes an event
horizon, which encircles the singularity, the region beyond which no
matter or light can escape. Mazur and Motolla \cite{3a} introduced
the concept of a gravastar, a stellar structure designed to explore
the issues of singularities and event horizons. A defining
characteristic of this novel compact object is its lack of a
singularity. To avoid the formation of a singularity, a de Sitter
interior is employed, while a thin layer of exotic matter separates
the inner and outer layers. Each region's properties are governed by
a particular equation of state (EoS). Some other interesting works
can be found in \cite{Jack2,Jack1,Jack3,Jack5,Jack4}.

While indirect evidence in the literature suggests the possible
existence of gravastars, no direct observational proof exists yet.
Sakai et al. \cite{3} proposed detecting gravastars through their
shadows. Kubo and Sakai \cite{4} also suggested that gravitational
lensing could help identifying them, noting that black holes lack
microlensing effects of maximum brightness. The detection of
GW150914 by LIGO interferometers \cite{5,6} hinted at ringdown
signals from sources without an event horizon. Additionally, a
gravastar-like shadow was observed in a recent M87 image from the
Event Horizon Telescope (EHT) collaboration \cite{7}. Several
studies on gravastars focus on various mathematical and scientific
challenges, mostly within GR. Bilic et al. \cite{10} replaced the de
Sitter interior with a Chaplygin gas EoS, treating the system as a
Born-Infeld phantom gravastar. Lobo \cite{11} used dark energy to
replace the interior vacuum. To resolve the singularity, Lobo and
Arellano \cite{12} connected the Schwarzschild exterior with
internal non-linear electrodynamic geometries. Ghosh et al.
\cite{14} stated that a 4-dimensional gravastar cannot be extended
to higher dimensions.

Gravastars have attracted significant attention from astrophysicists
seeking to understand their structural properties. Visser and
Wiltshire \cite{22a} identified stable configurations for specific
EoS parameters by analyzing the effects of radial perturbations on
gravastar stability. Horvat et al. \cite{17} studied the radial
stability of continuous pressure gravastars using eigenvalue
solutions of Einstein's equations, identifying a critical energy
density point that distinguishes stable from unstable
configurations. Rahaman et al. \cite{16} proposed a charge-free
gravastar model in anti-de Sitter BTZ spacetime, highlighting its
non-singular nature. Das et al. \cite{18} introduced a stellar model
with singularity-free solutions in alternative gravity. Sharif and
Naz \cite{19} explored the impact of charge on gravastar
characteristics within the framework of energy-momentum squared
gravity. Inspired by the studies \cite{zyousaf1,19a,19b}, we present
advanced solutions for three areas with distinct EoSs. The study of
gravastars has evolved through different metric ansatz, with one of
the key metrics being the Kuchowicz metric spacetime \cite{15}.
Early works utilized this non-singular metric potential to study
gravastar characteristics \cite{15a}, and recent contributions
\cite{20,23a,21,22,23} have continued to examine gravastars within
this framework. These studies together provide a comprehensive
understanding of gravastars, incorporating both standard GR and
modified gravity theories. Some other interesting works on compact
stars can be found in
\cite{23d,23e,23c,23x,23y,23z,ts0,ts1,ts2,ts3,ts4,ts5,ts6,tk1,tsl1,tsl5,tsl6,tandrade1,tandrade2,
tandrade3,zyousaf4,zyousaf5,ti1,ti2,ti3,ti4,ti5,ti6,yousaf4,yousaf5,yousaf6,23b,23f,tattiq1,tattiq2,
thira1,twaseem3,ti7}.

In this paper, we use the Kuchowicz metric to investigate the
gravastar geometry within the framework of Rastall theory. We
explore the graphical behavior of various gravastar properties for a
model corresponding to the intrinsic shell. The structure of the
paper is given as follows. Section \textbf{2} introduces the basic
formalism of the modified FEs with the temporal Kuchowicz metric
component. Section \textbf{3} examines the gravastar geometry across
three distinct regions, each with its respective EoS, while ensuring
the modification of the junction conditions in Rastall theory. In
section \textbf{4}, we examine the boundary conditions, match the
interior and thin shell solutions at the inner boundary, and the
thin shell and exterior solutions at the outer boundary, to
determine the required constants. Section \textbf{5} investigates
several fundamental attributes of the gravastar thin shell. Finally,
section \textbf{6} concludes the paper with a summary of the key
findings within the non-conservative gravity.

\section{Geometry of Rastall Field Equations}

In contrast to Einstein's GR, Rastall theory relaxes the standard
energy-momentum conservation law, allowing its covariant divergence
to be non-zero in curved spacetime, leading to
\begin{equation}\label{q1}
\nabla_{\theta} T^{\theta\delta} = \mu\mathbf{R}^{;\delta},
\end{equation}
with $\mathbf{R}$ being the Ricci scalar and $\mu$ is the Rastall
coupling parameter, which measures the deviation of Rastall theory
from GR. Unlike GR, the standard conservation law of the EMT is only
recovered in Minkowski spacetime. This arises naturally due to an
explicit coupling between matter and geometry introduced via $\mu$.
Hence, in Rastall theory, a non-flat spacetime geometry is
necessitated. Imposing the modified conservation condition leads to
the following modified FEs as
\begin{equation}\label{q2}
G_{\theta\delta} + \kappa \mu\, g_{\theta\delta} \mathbf{R} = \kappa
T_{\theta\delta}.
\end{equation}
Here, \( \kappa \) refer to the gravitational coupling constant.
Rastall \cite{4a} demonstrated that Eq.\eqref{q2} leads to the
following relation when taking the trace as
\begin{equation}\label{q3}
\mathbf{R}(4\kappa\mu - 1) = T,
\end{equation}
which implies that the trace of the EMT is generally non-zero.
Moreover, by considering the Newtonian limit and introducing
the dimensionless Rastall parameter \( \varkappa = \kappa \mu \),
one can express both \( \kappa \) and \(\mu \) as functions of
\(\varkappa \) as follows
\begin{equation}\label{3}
\kappa = \frac{8\pi G}{c^{4}} \frac{(4\varkappa - 1)}{(6\varkappa -
1)}, \qquad \mu=\frac{c^{4}}{8\pi G}\frac{\varkappa(6\varkappa- 1)}
{(4\varkappa - 1)}. \tag{3}
\end{equation}
Equation \eqref{3} explicitly defines the constant $\kappa$ and the
parameter $\mu$ in terms of the Rastall parameter $\varkappa$. These
expressions show that the standard Einstein value \( \kappa = 8\pi
\) is recovered in the limit \( \mu = 0 \), which corresponds to
\(\varkappa = 0 \). Notably, \( \kappa \) becomes singular when
\(\varkappa = 1/6 \), making this value physically inadmissible.
Furthermore, as seen in Eq.\eqref{3}, \(\mu \) becomes infinite when
\(\varkappa = 1/4\), indicating that the case \(\varkappa = 1/4 \)
is also not permitted. Therefore, the Newtonian limit reveals that
both these values of \(\varkappa \) are allowed in order to get
physically valid results. Finally, the FEs in Rastall theory takes
the form
\begin{equation}\label{q4}
G_{\theta\delta} +  \varkappa g_{\theta\delta} \, \mathbf{R} =
\kappa T_{\theta\delta} \frac{(4\varkappa - 1)}{(6\varkappa - 1)}.
\end{equation}
To characterize the interior spacetime of a static, spherically
symmetric object, we use the following line element of a
4-dimensional spacetime expressed in Schwarzschild coordinates \(
(x^i = t, r, \theta, \phi) \) as
\begin{equation}\label{q4a}
ds^2 = -e^{\omega(r)} dt^2 + e^{\psi(r)} dr^2 +r^2 d\theta^2 +  r^2
\sin^2 \theta \, d\phi^2,
\end{equation}
where \(\omega\) and \(\psi\) are functions depending only on the
radial coordinate \( r \). We now consider that the matter
distribution inside the star is that of a perfect fluid, which can
be expressed by the EMT as follows
\begin{equation}\label{q4b}
T_{\theta\delta} = (\rho + P) \eta_\theta
\eta_\delta+ P\, g_{\theta\delta},
\end{equation}
where \(\rho\) and \(P\) denote the energy density and isotropic
pressure, and \(\eta^\theta\) is the fluid's 4-velocity, satisfying
the relation \(\eta^\theta \eta_\theta = -1\). Hence, within the
context of non-conserved theory, using the spacetime metric
\eqref{q4a} along with the EMT \eqref{q4b}, the modified FEs as
expressed in Eq.\eqref{q4} take the form
\begin{align}
\bigg(\frac{4\varkappa - 1}{6\varkappa- 1}\bigg)\kappa \rho(r)&=
e^{-\psi} \left( \frac{\psi'}{r} - \frac{1}{r^{2}} \right) +
\frac{1}{r^{2}} +\varkappa e^{-\psi} \bigg\{ \omega^{\prime\prime} +
(\omega^{\prime})^2 + \psi^{\prime} \omega^{\prime}\notag
\\\label{q5}
&\quad-\frac{2}{r}(\psi^{\prime}-\omega^{\prime})-
\frac{2(e^{\psi} - 1)}{r^2}\bigg\}, \\
\bigg(\frac{4\varkappa-1}{6\varkappa-1}\bigg)\kappa P(r)&= e^{-\psi}
\left( \frac{\omega'}{r} + \frac{1}{r^{2}} \right) - \frac{1}{r^{2}}
- \varkappa e^{-\psi} \bigg\{ \omega^{\prime\prime} +
(\omega^{\prime})^{2} + \psi^{\prime}
\omega^{\prime}\notag\\\label{q6}
&\quad-\frac{2}{r}(\psi'-\omega')-\frac{2(e^{\psi}-1)}
{r^{2}}\bigg\},\\\nonumber
\left(\frac{4\varkappa-1}{6\varkappa-1}\right)\kappa P(r)
&=e^{-\psi} \left( \frac{\omega''}{2} - \frac{\psi'\omega}{4} +
\frac{\omega'^{2}}{4} + \frac{\omega' - \psi'}{2r} \right)
-\varkappa e^{-\psi} \notag \\\label{q7} &\times\left\{
\omega^{\prime\prime} + (\omega^{\prime})^2 + \psi^{\prime}
\omega^{\prime} - \frac{2}{r} (\psi^{\prime} -
\omega^{\prime})-\frac{2 (e^{\psi}-1)}{r^2} \right\},
\end{align}
where $'$ denotes $\frac{d}{dr}$. The modified conservation law in
Rastall framework, which balances pressure gradients, gravitational
pull and the non-conservative Rastall term, takes the form
\begin{equation}\label{q8}
\frac{dP}{dr} + \frac{1}{2} \frac{d\omega}{dr}( \rho+P) -
\frac{\varkappa}{4\varkappa - 1} \left(\frac{d\rho}{dr}-
3\frac{dP}{dr} \right) = 0.
\end{equation}
This generalized Tolman-Oppenheimer-Volkoff equation governs the
equilibrium of the gravastar shell, and by solving it, one can
verify whether the proposed configuration satisfies both hydrostatic
balance and stability criteria. In the limit $\varkappa\to0$, the
Rastall corrections vanish and Eq.\eqref{q8} smoothly reduces to the
standard equation of GR. The literature also presents some
fascinating works in different fields
\cite{Jack8,Jack6,Jack7,Jack9,Jack10}.

\section{Gravastar Model}

In this section, we derive the solution to FEs for a gravastar model
and analyze their physical and geometrical implications within
Rastall theory. The structure of a gravastar comprises
\begin{itemize}
\item \textbf{Core ($0 \leq r < r_1 = \mathbb{R}$):} A de
Sitter interior with EoS $\rho = -P$.
\item \textbf{Shell ($r_1= \mathbb{R} \leq r \leq r_2 = \mathbb{R} +
\epsilon$):} A thin, ultra-stiff layer admitting EoS $\rho = P$.
\item \textbf{Exterior ($r_2 < r =\infty$):} Vacuum region
described by the spherically symmetric spacetime with EoS $\rho = P
= 0$.
\end{itemize}
Here, $r_1$ and $r_2$ mark the shell boundaries and $\epsilon =
r_2 - r_1$ denotes its small thickness.

\subsection{Inner Domain}

The gravastar core is modeled as a de Sitter characterized by the
EoS given as
\[P + \rho = 0.\]
The negative pressure in this region produces a repulsive force that
counterbalances the inward gravitational attraction exerted by the
surrounding shell. This negative energy density effectively mimics a
positive cosmological constant. Such an EoS, often termed a
\textit{degenerate vacuum}, is commonly employed to model
dark energy. Substituting this EoS into the modified conservation
equation \eqref{q8} leads to
\begin{equation} \label{a1}
P = -\rho = -\rho_c,
\end{equation}
where \( \rho_c \) is a matter density constant, suggesting uniform
pressure and energy density throughout the interior region.
Utilizing this relation within Eqs.\eqref{q5}-\eqref{q7}, the metric
potential \( \psi(r) \) can be derived as follows
\begin{equation}\label{a2}
e^{-\psi(r)} = \frac{8\pi \rho_c r^{2}}{3} \left(1 + \frac{4\varkappa - 1}
{4\varkappa} \right) - \frac{\mathcal{X}_1}{3r} + 1,
\end{equation}
where \( \mathcal{X}_1 \) is an integration constant. To ensure
regularity at the center (\( r = 0 \)), we set \( \mathcal{X}_1 = 0
\), resulting in a simplified and regular solution as
\begin{equation} \label{a3}
e^{-\psi(r)} = \frac{8\pi \rho_c r^{2}}{3} \left(1 + \frac{4\varkappa - 1}
{4\varkappa} \right) + 1.
\end{equation}
The relationship between the metric functions \(\psi(r)\) and \(
\omega(r) \), using Eqs.\eqref{q5}, \eqref{q6} and EoS \eqref{a1},
is given by
\begin{equation} \label{a4}
e^{\omega(r)} = \mathcal{X}_2 e^{-\psi(r)},
\end{equation}
where \( \mathcal{X}_2\) is another constant. Since \( \rho_c \) is
density constant across the inner domain, the gravitational mass
function contained within a radius \(r_1 = \mathbb{R}\) is given by
\begin{equation} \label{a5}
\mathbf{M}(\mathbb{R}) = \int_0^{\mathbb{R}} 4\pi r^2 \rho\,
dr = \frac{4}{3} \pi \mathbb{R}^3 \rho_c,
\end{equation}
which represents the total mass enclosed within the interior region.
Figure \textbf{1} shows a monotonically increasing active mass
profile, confirming the lack of singularities and indicating a
smoothly distributed matter configuration within the gravastar's
inner region.
\begin{figure}\centering
\epsfig{file=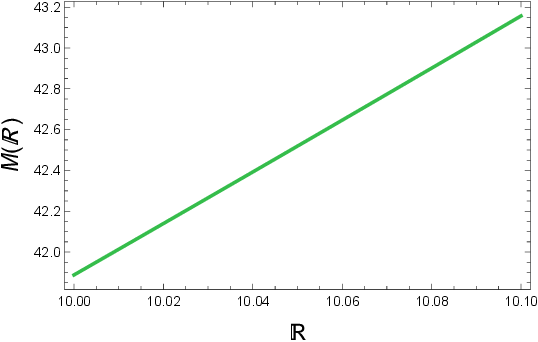,width=0.5\linewidth} \caption{Variation of
gravitational mass \eqref{a5} with radius $\mathbb{R}$ for $\rho_{c}
= 0.25$.}
\end{figure}

\subsection{Intermediate Shell Domain}

The gravastars interior is enclosed by a thin shell of
ultra-relativistic fluid (or ``soft quanta''), which satisfies the
stiff-fluid EoS given as
\begin{equation}\label{17}
P = \rho.
\end{equation}
This was originally introduced by Zel\'dovich to model a cold,
baryonic universe \cite{6e}. This stiff-fluid EoS has since been
adopted by many cosmologists and astrophysicists
\cite{44e,45e,46e,47e,48e}. Solving the Einstein FEs exactly within
this non-vacuum shell is intractable in general. Therefore, we adopt
the approximation as $0 < e^{-b(r)} \ll 1$ in the ultra-relativistic
thin shell. Physically, the matching of two distinct spacetimes
requires an intermediate layer of negligible thickness \cite{49f}.
Moreover, because the shell is ultra-thin, any quantity that depends
on \(r\) can be treated as vanishingly small as\(r \to 0\), enabling
us to obtain the following equations from \eqref{q5}-\eqref{q7} as
\begin{align}\label{20}
\left(\frac{4\varkappa-1}{6\varkappa-1}\right)\rho(r) &= \frac{e^{-\psi}
\psi'}{r}+\frac{1}{r^2} - \frac{2\varkappa}{r^2}+ e^{-\psi}\varkappa
\left(\psi'\omega'+\frac{2\psi'}{r}\right),
\\\label{21}
\left(\frac{4\varkappa-1}{6\varkappa-1}\right)P(r) &=\frac
{2\varkappa}{r^2}-\frac{1}{r^2}+e^{-\psi}\varkappa
\left(\psi'\omega'+ \frac{2\psi'}{r}\right),\\\label{22}
\left(\frac{4\varkappa-1}{6\varkappa-1}\right)P(r) &=\frac{2
\varkappa}{r^2}+\frac{e^{-\psi}\psi'}{r}\left(\omega'+2\right)
-\frac{e^{-\psi}}{4}\bigg(\psi'\omega'+\frac{2\psi'}{r}\bigg).
\end{align}
In studying stable gravastar configurations within Rastall theory,
selecting a suitable metric potential is essential to obtain
physically consistent solutions to the modified FEs. We employ the
Kuchowicz potential to represent the temporal component of the
interior spacetime metric in an intermediate thin shell. This metric
ansatz, characterized by few adjustable constants, yields a
non-singular and horizon-free structure, yields stable solutions
consistent with the three-layer gravastar model. This non-singular
metric function, originally proposed by Kuchowicz, has been
effectively used to investigate stable celestial configurations in
various gravitational theories. We assume the metric function \(
e^{\omega(r)}\) to take the form proposed by Kuchowicz \cite{15},
expressed as
\begin{equation}\label{12}
e^{\omega(r)} = \mathbf{C}^2 e^{\mathbf{B} r^{2}},
\end{equation}
where \(\mathbf{B}\) is an unknown constant with dimensions of \(
\text{length}^{-2} \), and \(\mathbf{C}\) is a dimensionless
parameter. Putting Eq.\eqref{12} into \eqref{20}-\eqref{22} leads to
\begin{align}
\label{20a}
\left(\frac{4\varkappa-1}{6\varkappa-1}\right)\rho(r) &=
\frac{e^{-\psi} \psi'}{r} + \frac{1}{r^2} - \frac{2\varkappa}{r^2} +
\varkappa e^{-\psi}\, \psi' \left(2\mathbf{B} r + \frac{2}{r}\right), \\
\label{21b}
\left(\frac{4\varkappa-1}{6\varkappa-1}\right)P(r) &=
\frac{2\varkappa}{r^2} - \frac{1}{r^2} + \varkappa e^{-\psi}\,
\psi' \left(2\mathbf{B} r + \frac{2}{r}\right), \\
\label{22c}
\left(\frac{4\varkappa-1}{6\varkappa-1}\right)P(r) &=
\frac{2\varkappa}{r^2} + \frac{e^{-\psi} \psi'}{r}
(2\mathbf{B} r + 2) - \frac{e^{-\psi} \psi'}{4}
\left(2\mathbf{B} r + \frac{2}{r}\right).
\end{align}
By combining Eqs.\eqref{20a} and \eqref{21b} with the previously
stated EoS \eqref{17}, we obtain the inverse radial metric function
for the shell, given by
\begin{equation}\label{25a}
e^{-\psi(r)}=\frac{(2-4\varkappa ) \big[\ln \left (1-4\varkappa
\left(\mathbf{B} r^2+1\right)\right) -2 \ln (r)\big]}{8\varkappa
-2}+ \mathcal{X}_3.
\end{equation}
\begin{figure}\centering
\epsfig{file=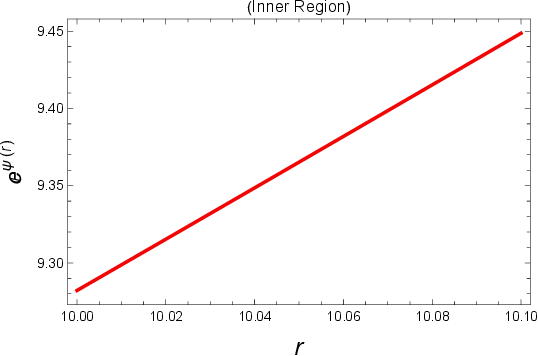,width=.5\linewidth}\epsfig{file=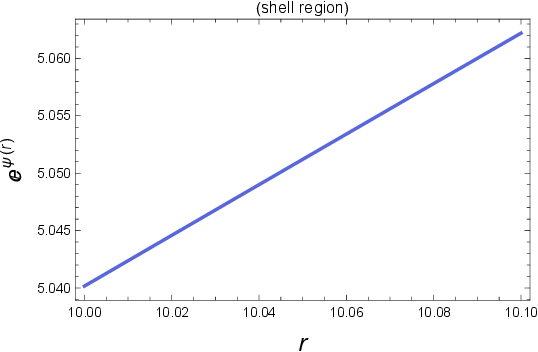,width=.5\linewidth}
\caption{Variation of metric potential $e^{\psi(r)}$ \eqref{a3} and
\eqref{25a} with radius $r$.}
\end{figure}
The integration constant, denoted as $\mathcal{X}_{3}$, can be
determined by applying the boundary conditions. The behavior of the
metric function \(e^{\psi}\) for inner and shell domain, shown in
Figure \textbf{2} with respect to the radial coordinate,
demonstrates that it remains regular and free of singularities
throughout the region of gravastar. By substituting Eq.\eqref{25a}
together with the stiff fluid EoS and the Kuchowicz metric function
into \eqref{q8}, the expression for the matter density= pressure is
obtained as
\begin{equation}\label{25b}
P = \rho = \rho_{c}e^{-\mathbf{B} r^{2}(\frac{4\varkappa-1}{6\varkappa-1})}.
\end{equation}
The variation in matter density across the shell is depicted in
Figure \textbf{3} that illustrates positive and decreasing nature of
the pressure=density throughout the shell, with a sharp decrease as
the radial distance increases.
\begin{figure}\centering
\epsfig{file=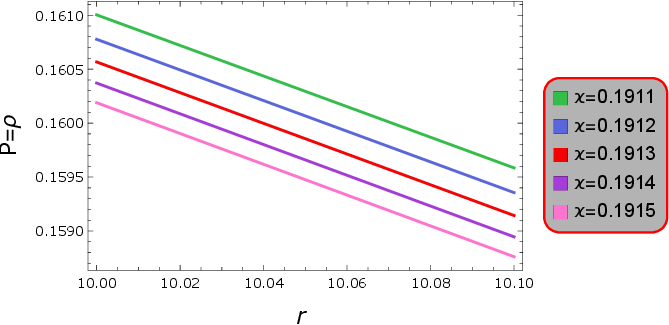,width=.6\linewidth}\caption{Variation of
pressure-density profile \eqref{25b} with radius $r$ for
$\rho_{c}=0.25$.}
\end{figure}

\subsection{Outer Domain and Israel Matching Constraints}

The exterior region \((r > r_2)\) of the gravastar is modeled by the
\((3+1)\)-dimensional Schwarzschild metric, representing the vacuum
spacetime around a static, spherically symmetric geometry. This
region satisfies Einstein's FEs with the EoS \(P = \rho = 0\). The
Schwarzschild metric is given by
\begin{equation}
\label{eq:schwarzschild}
ds^2 = -\left(1 - \frac{2M}{r}\right) dt^2 + \frac{1}{\bigg(1 -\frac{2M}{r}\bigg)}
dr^2 + r^2 d\theta^2 + r^2 \sin^2 \theta \, d\phi^2,
\end{equation}
where
\begin{itemize}
\item \(M\) is the total gravitational mass.
\item \( r > 2M \), lies outside the Schwarzschild radius.
\end{itemize}
Physically, the Schwarzschild solution describes the spacetime
generated by a spherically symmetric mass in the absence of matter
or radiation (i.e., in vacuum). For realistic astrophysical models,
it is essential to achieve a seamless transition between the
interior and exterior spacetimes, denoted by \(\mathbb{M}^{\pm}\),
across the hypersurface \(\Omega\).

Darmois \cite{47f} and Israel \cite{48f} formulated the junction (or
matching) conditions that require the continuity of the induced
metric across \(\Omega\). The extrinsic curvature may exhibit a
discontinuity at the junction radius \(r = \mathbb{R}\) gives rise
to a non-zero surface EMT, governed by the Lanczos equations
\cite{49f} given as
\begin{equation}
\mathbf{S}^{\gamma}_{\sigma} = -\frac{1}{8\pi} \left
(\chi^{\gamma}_{\sigma} - \delta^{\gamma}_{\sigma}
\, \chi^{a}_{a} \right),
\end{equation}
where \(\chi^{\gamma}_{\sigma}  =\chi^{+\,\gamma}_{\sigma} -
\chi^{-\,\gamma}_{\sigma}\) represents the discontinuity in the
extrinsic curvature across the junction surface. The extrinsic
curvature (also referred to as the second fundamental form) on each
side of the hypersurface, embedded in the respective manifolds
\(\mathbb{M}^{\pm}\), is defined as
\begin{equation}
\chi_{\sigma\gamma} = \frac{\partial x^{l}}{\partial
\zeta^{\sigma}} \frac{\partial x^{m}}{\partial\zeta^{\gamma}}
\nabla_{l}\varrho_{m},
\end{equation}
which can be expressed as
\begin{equation}
\chi^{\pm}_{\sigma\gamma} = -n^{\pm}_{m} \left[\frac{\partial^2
x^m}{\partial \zeta^{\sigma}\partial \zeta^{\gamma}} +
\Gamma^{m}_{pq} \frac{\partial x^{p}} {\partial
\zeta^{\sigma}}\frac{\partial x^{q}}{\partial \zeta^{\gamma}}
\right]_{\Omega}, \tag{33}
\end{equation}
where \(\zeta^{\sigma}\) are the coordinates on the shell (i.e., the
intrinsic coordinates of the junction surface) and \( n_{m}\) are
the unit normal vectors to the hypersurface $\Omega$ from interior
($-$) to the exterior ($+$), defined as
\[n^{\pm}_{m} = \pm \bigg|g^{pq} \frac{\partial
\mathcal{H}}{\partial x^{p}} \frac{\partial \mathcal{H}} {\partial
x^{q}}\bigg|^{-\frac{1}{2}}\frac{\partial \mathcal{H}}{\partial
x^{m}}, \quad n^{m}n_{m} = 1. \tag{34}\] By applying the Lanczos
equations, the surface EMT \( \mathbf{S}^{\sigma}_{\gamma} \) takes
the form, $\mathbf{S}^{\sigma}_{\gamma} = \mathrm{diag}(\Xi,\
-\Upsilon -\Upsilon,\ -\Upsilon),$ where \(\Xi\) represents the
surface energy density, and \(\Upsilon\) denotes the surface
pressure. The explicit expressions for \(\Xi\) and \(\Upsilon\) can
be computed as
\begin{equation}
\Xi = -\frac{1}{4\pi \mathbb{R}} \left[ \sqrt{\mathcal{H}} \right]^{+}_{-},
\qquad
\Upsilon = -\frac{\Xi}{2} + \frac{1}{16\pi} \left[ \frac{\mathcal{H}}
{\sqrt{\mathcal{H}}} \right]^{+}_{-}.
\tag{35}
\end{equation}
The notation \( [\mathcal{H}]^{+}_{-}\) denotes the jump of the
quantity \(\mathcal{H}\) across the shell, i.e., \( \mathcal{H}^{+}
- \mathcal{H}^{-} \). By substituting the inner and outer geometries
of the gravastar into the expressions above, one can obtain the
explicit forms of surface quantities as
\begin{align}\label{52}
\Xi&= \frac{1}{4\pi
\mathbb{R}}\bigg[\sqrt{\frac{8\pi\rho_{c}\mathbb{R}^2(1+\frac
{4\varkappa}{4\varkappa-1})}{3}+1}-\sqrt{1-\frac{2M}{\mathbb{R}}}\bigg],
\\\label{53} \Upsilon&= \frac{1}{8\pi
\mathbb{R}}\bigg[{\frac{1-\frac{M}{\mathbb{R}}}{\sqrt{1-\frac{2M}
{\mathbb{R}}}}-\frac{\frac{16\pi \rho_{c}}{3}{(1+\frac{4\varkappa}
{4\varkappa-1})\mathbb{R}^2}+1}{\sqrt{\frac{8\pi\rho_{c}
}{3}{(1+\frac {4\varkappa}{4\varkappa-1})\mathbb{R}^2}+1}}}\bigg].
\end{align}
The changes in surface energy density and surface pressure is shown
in Figure \textbf{4}. Both quantities admit positive and finite
nature which ensures the well-behaved and physically realistic
nature of the thin shell.
\begin{figure}[h!]\centering
\epsfig{file=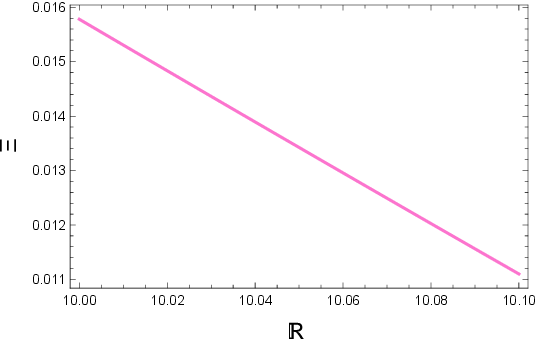,width=.5\linewidth}\epsfig{file=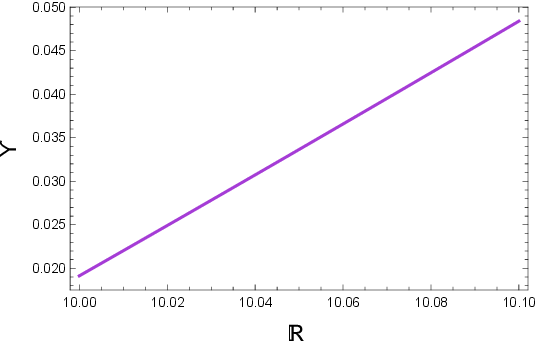,width=.5\linewidth}
\caption{Variation of surface energy density \eqref{52} and surface
pressure \eqref{53} with radius $\mathbb{R}$ for $\rho _c=0.25$.}
\end{figure}
The surface mass of the thin shell can now be calculated using the
surface pressure as
\begin{equation}\label{54}
m_\mathbf{shell}=4\pi \mathbb{R}^{2}\Upsilon =
\mathbb{R}\bigg[\sqrt{1+\frac{8\pi \rho_{c}\mathbb{R}^2}{3}
{\bigg(1+\frac{4\varkappa}{4\varkappa-1}\bigg)
}}-\sqrt{1-\frac{2M}{\mathbb{R}}}\bigg].
\end{equation}
The total mass $M$ of the gravastar, expressed in the terms of the
surface mass $m_\mathbf{shell}$, is calculated as
\begin{equation}\label{55}
M=\frac{1}{6 \mathbb{R}}\bigg[2 \mathbb{R}{m_\mathbf{shell}
\sqrt{24\pi \mathbb{R}^{2}+9}}-3m^{2}_\mathbf{shell}-
8\pi\mathbb{R}^{4}\bigg].
\end{equation}

\subsection{Equation of State Parameter}

For the gravastar's thin shell, Eqs.\eqref{52} and \eqref{53} lead
to the EoS parameter at \( r = \mathbb{R} \), resulting in
 \begin{equation}\label{w1}
\mathcal{W}(\mathbb{R})=\frac{\Upsilon}{\Xi}.
\end{equation}
Substituting the values for $\Xi$ and $\Upsilon$, we find
\begin{equation}\label{w2}
\mathcal{W}(\mathbb{R})=\frac{\bigg[{\frac{1-\frac{M}{\mathbb{R}}}
{\sqrt{1-\frac{2M}{\mathbb{R}}}}-\frac{\frac{16\pi \rho_{c}}{3}{(1+\frac
{4\varkappa}{4\varkappa-1})\mathbb{R}^2}+1}{\sqrt{\frac{8\pi\rho_{c}}
{3}{(1+\frac{4\varkappa}{4\varkappa-1})\mathbb{R}^2}+1}}}\bigg]}{2\bigg
[\sqrt{\frac{8\pi\rho_{c}\mathbb{R}^2(1+\frac{4\varkappa}{4\varkappa-1})}
{3}+1}-\sqrt{1-\frac{2M}{\mathbb{R}}}\bigg]}.
\end{equation}
The positive matter density and pressure ensure a positive value for
\( \mathcal{W}(\mathbb{R}) \). To guarantee that the EoS parameter
remains real, the following conditions must hold: \(
\frac{2M}{\mathbb{R}} < 1 \) and \( \frac{8\pi \rho_c \mathbb{R}^2
\left( 1 + \frac{4\varkappa}{4\varkappa - 1} \right)}{3} < 1 \).
Expanding the square root terms in both the numerator and
denominator of Eq.\eqref{w2} as a binomial series, applying \(
\frac{M} {\mathbb{R}}\ll 1 \) and \( \frac{8\pi \rho_c \mathbb{R}^2
\left( 1 +\frac{4\varkappa} {4\varkappa - 1} \right)}{3} \ll 1 \),
and retaining first-order terms, we attain
\begin{equation}\label{w2}
\mathcal{W}(\mathbb{R}) \approx \frac{3}{2 \left[ \frac{3M}{4\pi \rho_c
\mathbb{R}^3 \left( 1 + \frac{4\varkappa}{4\varkappa - 1} \right)} - 1 \right]}.
\end{equation}
For \( \mathcal{W}(\mathbb{R}) \) to be positive, the denominator
must be positive, which implies the condition: $\frac{3M}{4\pi
\rho_c\mathbb{R}^3\left(1+\frac{4\varkappa} {4\varkappa - 1}
\right)} > 1$. Violation of this condition may result in a negative
or undefined value for \( \mathcal{W}(\mathbb{R}) \). Figure
\textbf{5} shows that, as the shell radius increases, the parameter
grows, indicating a stiffer EoS with higher pressure for a given
energy density as the shell expands.
\begin{figure}\centering
\epsfig{file=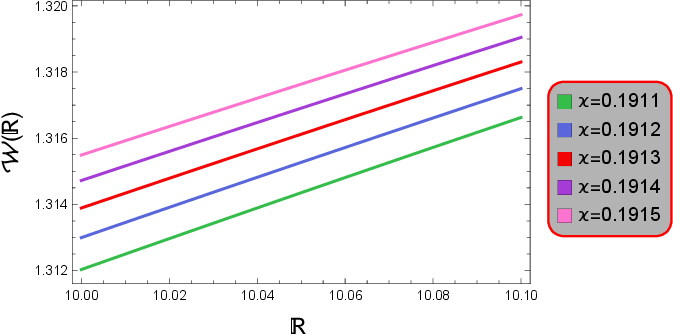,width=.6\linewidth}\caption{Variation of EoS
parameter \eqref{w2} with radius $\mathbb{R}$ for $\rho_c=0.25 $ .}
\end{figure}

\section{Boundary Conditions at Interface}

For a gravastar to remain stable, the spacetime metric must be
continuous across both key interfaces: the interior-shell boundary
$(r=r_{1})$ and the shell-exterior boundary $(r=r_{2})$. Ensuring
continuity at $\mathbb{R}= \mathbb{R}_2$ (linking the shell to the
exterior spacetime in surface boundary) requires not only the
matching of the metric components $g_{tt}$ and $g_{rr}$ but also the
continuity of the radial derivative component $\frac{\partial
g_{tt}} {\partial r}$. By matching the metric functions at these
interfaces, we can solve for the unknown constants $\mathbf{B}$,
$\mathbf{C}$, and $\mathcal{X}_{2}$ involved in the analysis. This
leads to a system of three independent equations as
\begin{align}
\frac{\partial g_{tt}}{\partial r}: \quad & \frac{2 M}{\mathbb{R}_2^2}
=\frac{2 \mathbf{B} \mathbb{R}_2 e^{\mathbf{B} \mathbb{R}_2^2}}
{\mathbf{C}^2},\tag{21}\\
g_{tt}: \quad & 1 - \frac{2M}{\mathbb{R}_2}=e^{\mathbf{B}
\mathbb{R}_2^2-2 \ln (\mathbf{C})},\tag{20} \\
g_{rr}: \quad & 1-\frac{2M}{\mathbb{R}_2}=\frac{(2-4\varkappa )
\left(\ln \left(1-4\varkappa\left(\mathbf{B}
\mathbb{R}_2^2+1\right)\right)-2 \ln
\left(\mathbb{R}_2\right)\right)}{8\varkappa -2}+\mathcal{X}_{2},
\tag{22}
\end{align}
whose simultaneous solution gives
\begin{align}
\mathbf{B} &= -\frac{M}{\mathbb{R}_2^2 \left(2 M-\mathbb{R}_2\right)},\tag{23} \\
\mathbf{C} &=\sqrt{e^{-\frac{M}{2 M-\mathbb{R}_2}}}\sqrt{-\frac
{\mathbb{R}_2}{2 M-\mathbb{R}_2}},\tag{24}\\
\mathcal{X}_{2} &=-\frac{(2-4\varkappa) \left(\ln
\left(1-4\varkappa\left(\mathbf{B} \mathbb{R}_2^2+1\right)\right)-2
\ln \left(\mathbb{R}_2\right)\right)} {8\varkappa
-2}-\frac{2M}{\mathbb{R}_2}+1.\tag{25}
\end{align}
This matching of metric functions guarantees a smooth spacetime
geometry without physical or geometric singularities at the
interfaces. For the numerical evaluation of the gravastar's physical
properties, we select a specific value for the Rastall parameter
$\varkappa$ and substitute it along with other key quantities.
Specifically, we set the gravastar mass to $M = 3.75\,M_{\odot}$,
the interior radius to $\mathbb{R}_1 = 10.00\,km$, and the exterior
radius to $\mathbb{R}_2 = 10.10\, km$. Additionally, the central
density is fixed at $\rho_c = 0.25$, to study how various physical
quantities vary with respect to these inputs. With these choices, we
compute the constants $\mathbf{B}$, $\mathbf{C}$ and
$\mathcal{X}_2$, whose values are listed in Table \textbf{1}.
\begin{table}[H]
  \centering
  \scriptsize
  \caption{Values of constants for $M=3.75\,M_{\odot}$ and
  $\mathbb{R}_{2}=10.10\ km$.}
  \label{tab:constants}
  \setlength{\tabcolsep}{2.1em}
  \begin{tabular}{@{}ccccc@{}}
    \toprule
    $\varkappa$ & $M/\mathbb{R}$ & $\mathbf{B}$ &
    $\mathbf{C}$ & $\mathcal{X}_2$ \\
    \midrule
    0.1911 & 8.9778 & 0.035009 & 17.0165 & -9.60851    \\
    0.1912 & 8.9778  & 0.035009 & 17.0165 & -9.61998  \\
    0.1913 & 8.9778  & 0.035009 & 17.0165 & -9.6315    \\
    0.1914 & 8.9778  & 0.035009 & 17.0165 & -9.64305   \\
    \bottomrule
\end{tabular}
\end{table}

This specific case study leads to two key questions about the
parametric selection in our solution:
\begin{enumerate}
\item \textbf{Existence}: For any given \( M \) and \( \mathbb{R}_2 \),
does a regular (non-singular) solution always exist?
\item \textbf{Uniqueness}: Once \( M \) and \( \mathbb{R}_2 \) are fixed, is the
gravastar configuration unique, or can different values lead to distinct solutions?
\end{enumerate}
In this work, we select parametric values that both illustrate the
gravastar's physical behavior and satisfy the compactness
constraint. As long as this condition holds, any pair \( (M,
\mathbb{R}_2) \) will yield a solution with the same qualitative
features as those presented in this study.

\section{Physical Key Properties of Shell Domain}

This section examines the impact of the modified theory on the
physical features of the shell region. We calculate the shell's
proper length, total energy of the relativistic configuration, and
the entropy. Importantly, these quantities are derived without
relying on the thin-shell approximation, providing an exact
description of the shell's structure. These results will be
presented alongside the dynamical formulation of gravastars and
visualized through corresponding plots.

\subsection{Proper Thickness}

The gravastar's thin shell divides its region I and III, with the
inner boundary at $r_1 = \mathbb{R}$ and the outer boundary at $r_2
= \mathbb{R} + \epsilon$. Here, $\epsilon$ denotes the shell's
thickness, which is taken to be very small ($\epsilon \ll 1$),
indicating only a minimal change in radial distance across the
shell. The shell's proper thickness, accounting for spacetime
curvature, is given by the integral
\begin{equation}\label{z1}
\mathcal{L}=\int_{\mathbb{R}}^{\mathbb{R}+\epsilon}
\sqrt{ e^{\psi(r)}}dr,
\end{equation}
where \(g_{rr}(r) = e^{\psi(r)}\) is the radial metric component in
the shell. Substituting its value from Eq.\eqref{z1} yields
\begin{equation}\label{z2}
\mathcal{L}= \int_{\mathbb{R}}^{\mathbb{R}+\epsilon}\frac{1}
{\sqrt{\displaystyle\frac{(2 - 4\varkappa)\Bigl(\ln\bigl[1 - 4
\varkappa(\mathbf{B}r^2 + 1)\bigr] - 2\ln(r)\Bigr)}{8\varkappa - 2}
+ \mathcal{X}_2}}dr.
\end{equation}
To solve the above equation, we assume $\frac{d
H(r)}{dr}=\frac{1}{H(r)}$ with $H(r)=\sqrt{e^{\psi(r)}}$, so that
\begin{equation}\label{z5}
\mathcal{L}=\int_{\mathbb{R}}^{\mathbb{R}+\epsilon} \frac{dH(r)}{dr}dr
=H\bigl(\mathbb{R} + \epsilon\bigr)-H(\mathbb{R})
\approx\epsilon \left.\frac{dH(r)}{dr}\right|_{r=\mathbb{R}}.
\end{equation}
Equation \eqref{z2} thus becomes under the above result as
\begin{equation}\label{z6}
\mathcal{L} = \epsilon\sqrt{\frac{1}{\displaystyle \frac{(2 -
4\varkappa) \Bigl(\ln\bigl[1 - 4\varkappa(\mathbf{B}\mathbb{R}^2 +
1)\bigr] -2\ln(\mathbb{R})\Bigr)}{8\varkappa - 2}+\mathcal{X}_2}}.
\end{equation}
In this scenario, \( \epsilon \) is a small positive value, making
higher-order terms like \( \epsilon^2 \) negligible. Equation
\eqref{z6} establishes a direct relationship between the shell's
coordinate thickness and its proper length, explicitly depending on
the shell radius and other model parameters. Figure \textbf{6}
illustrates a consistent trend with increasing shell thickness,
demonstrating a uniform pattern across all evaluated physical
Rastall parameters.
\begin{figure}\centering
\epsfig{file=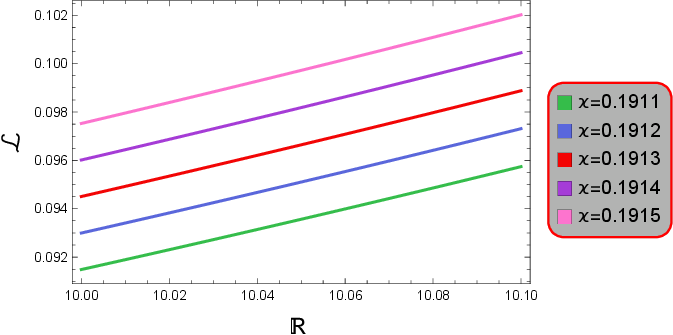,width=.6\linewidth}\caption{ Variation of Proper
length \eqref{z6} with radius $\mathbb{R}$ for $\epsilon=0.01$.}
\end{figure}

\subsection{Energy Content}

Inside the gravastar, the matter obeys the EoS $P = -\rho$,
representing a region with negative energy density. This
negative pressure gives rise to a repulsive force that opposes
gravitational collapse, thereby playing a crucial role to stabilize
the gravastar structure by preventing singularity formation.
The shell's total energy can be expressed as
\begin{equation}\label{s1}
\mathcal{E} = 4\pi \int_{\mathbb{R}}^{\mathbb{R}+\epsilon}
\rho\, r^2\, dr.
\end{equation}
Substituting the expression of \(\rho\) from Eq.\eqref{25b}, we
obtain
\begin{align}\label{s2}
\mathcal{E} =\left[\frac{\pi  (6 \varkappa -1) \rho _c \left(
\sqrt{6\pi\varkappa -\pi} \text{erf}\left(\frac{\sqrt{\mathbf{B}}
r \sqrt{4 \varkappa -1}}{\sqrt{6 \varkappa -1}}\right)-2
\sqrt{\mathbf{B}} r \sqrt{4 \varkappa -1} e^{\frac{\mathbf{B}r^2
(1-4 \varkappa )}{6 \varkappa -1}}\right)}{\mathbf{B}^{3/2}
(4\varkappa -1)^{3/2}}\right]_{\mathbb{R}}^{\mathbb{R}+\epsilon}.
\end{align}
The graph in Figure \textbf{7} shows a linear increase in energy
content as the thickness and radius of the shell expand, resulting
in a higher matter density and possibly more dense configurations.
\begin{figure}\centering
\epsfig{file=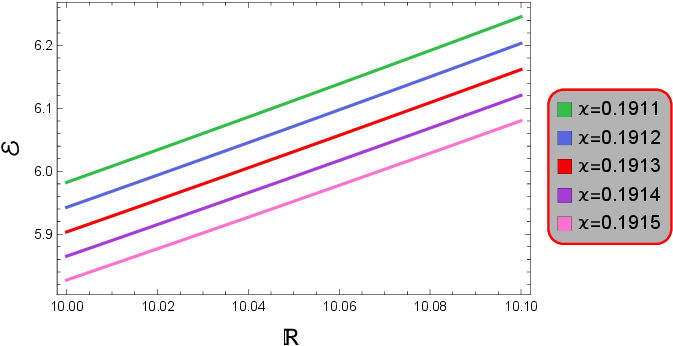,width=.6\linewidth}\caption{Variation of energy
\eqref{s2} with radius $\mathbb{R}$ for $\rho _c=0.25$.}
\end{figure}

\subsection{Entropy}

The entropy of a gravastar arises entirely from its shell, as there
is no event horizon. It is computed by integrating the local entropy
density over the shell's volume, taking into account both the
spacetime curvature and the shell's geometry. The typical formula
for the entropy $\mathcal{S}$ of the shell is expressed as
\begin{equation}\label{43} \mathcal{S} =  4\pi
\int_{\mathbb{R}}^{\mathbb{R}+\epsilon}
r^2\varrho(r)\sqrt{e^{\psi(r)}} dr,
\end{equation}
where \(\varrho(r)\) represents the entropy density, which is
defined in terms of the local temperature \(\tau(r)\) as
\begin{equation}\label{44}
\varrho(r) = \frac{\sigma^2 \mathcal{K}_B^2 \tau(r)}{4\pi \hbar^2}
=\frac{\sigma \mathcal{K}_B}{\hbar}\bigg(\frac{P}{2\pi}\bigg)^{1/2}.
\end{equation}
Here, \(\sigma\) represents a constant with no units. For
convenience, we adopt Planck units by setting \(\mathcal{K}_B = 1\)
and \(\hbar = 1\). Using the thin-shell approximation and applying
Taylor expansion, following a method similar to that used in the
proper length calculation \cite{19a}, the entropy \( \mathcal{S} \)
can be expressed as
\begin{align} \label{45}
\mathcal{S} &= \bigg[\epsilon(\mathcal{Z}) - \frac{1}{2 \mathbb{R}
\mathcal{Z} \mathcal{Z}_1 (6 \varkappa - 1) \left( (2 \varkappa - 1)
\left(2 \ln(\mathbb{R}) - \ln(\mathcal{Z}_1)\right)
+ \mathcal{X}_2 (4 \varkappa - 1) \right)^2 } \nonumber \\
&\quad \times \bigg\{\rho_c (\epsilon - 4 \epsilon \varkappa)^2
e^{\frac{\mathbf{B} \mathbb{R}^2 (1 - 4 \varkappa)}{6 \varkappa -
1}} \big(\mathbf{B}\mathbb{R}^2\mathcal{X}_2\mathcal{Z}_1
(4\varkappa - 1)\ln(e) +(2 \varkappa - 1) \nonumber \\
&\quad \times \left(- \mathbf{B} \mathbb{R}^2 \mathcal{Z}_1 \ln(e)
\left(2 \ln(\mathbb{R}) - \ln(\mathcal{Z}_1)\right) + 6 \varkappa -
1 \right) \big)\bigg\}\bigg]2 \sqrt{2 \pi}\sigma \mathbb{R}^2,
\end{align}
where the auxiliary functions \(\mathcal{Z}\) and
\(\mathcal{Z}_1\) are defined as
\begin{equation} \nonumber
\mathcal{Z} = \frac{\rho_c e^{\frac{\mathbf{B} \mathbb{R}^2 (1 -
4\varkappa)}{6\varkappa}}} {\frac{(2 - 4\varkappa)\left( \ln\left(1
- 4\varkappa(\mathbf{B} \mathbb{R}^2 + 1)\right)-2\ln(\mathbb{R})
\right)}{8\varkappa-2} + \mathcal{X}_2},\qquad \mathcal{Z}_1 = 1 -
4\varkappa\left(\mathbf{B} \mathbb{R}^2 + 1\right).
\end{equation}
The graphical depiction of entropy is shown in Figure \textbf{8}.
The total entropy increases with the thickness of the shell, as a
thicker shell provides more volume to store entropy. In contrast, if
the shell were infinitely thin, it would contribute negligibly to
the overall entropy.
\begin{figure}\centering
\epsfig{file=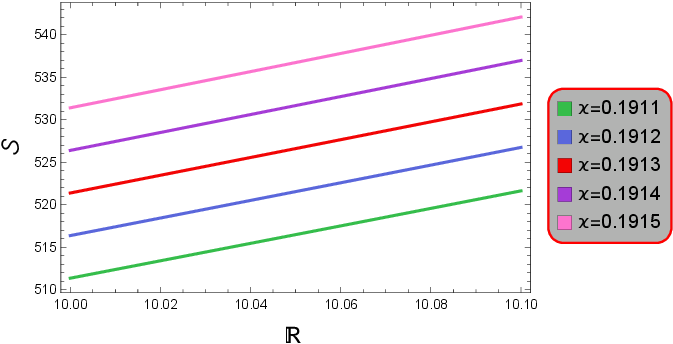,width=.6\linewidth}\caption{Variation of entropy
\eqref{45} with radius $\mathbb{R}$ for $\rho _c=0.25$.}
\end{figure}

\subsection{Stellar Stability Criteria}

Evaluating the stability of gravastar configurations is essential to
determine their potential as astrophysical entities. This subsection
will investigate two unique methods for assessing stability: surface
redshift analysis and the adiabatic index approach. By exploring
these factors, we can better understand whether gravastars can
maintain a stable state and how they may respond under varying
conditions.

\subsubsection{Surface Redshift}

Investigating the surface redshift of a gravastar is a fundamental
method for evaluating its stability and detectability. The
gravitational surface redshift, is defined as
$\mathcal{Z}_s=\frac{\Delta\Lambda}{\Lambda_e}=
 \frac{\Lambda_0}{\Lambda_e},$ where \(\Lambda_0 \) is the
observed wavelength and \(\Lambda_e\) is the emitted wavelength.
Buchdahl \cite{83z} showed that for a stable, isotropic perfect
fluid, the maximum surface redshift is 2. Ivanov \cite{59z} argued
that for anisotropic fluids, this limit can increase to 3.84.
Barraco and Hamity \cite{60z} demonstrated that \( \mathcal{Z}_s
\leq 2 \) holds for isotropic fluids in the absence of a
cosmological constant, while B\"{o}hmer and Harko \cite{50z} found
that with both anisotropy and a cosmological constant, the limit
rises to 5. In our analysis, the surface redshift is calculated
using the following expression
\begin{equation}\label{5r}
\mathcal{Z}_s=-1+\frac{1}{\sqrt{g_{tt}}}=-1+\frac{1}{\mathbf{C}
e^{\frac{\mathbf{B}r^{2}}{2}}}.
\end{equation}
We numerically solve Eq.\eqref{5r} and present the outcome in Figure
\textbf{9} which illustrates that the surface redshift is positive
for positive values of $\mathbf{B}$ and $\mathbf{C}$. This indicates
that our current investigation for gravastars is both stable and
physically consistent.
\begin{figure}
\epsfig{file=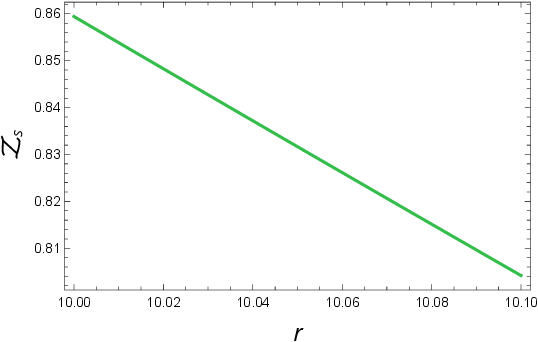,width=.5\linewidth}\epsfig{file=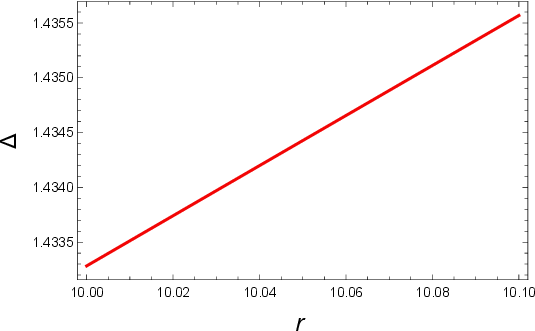,width=.5\linewidth}
\caption{Variation of the surface redshift $\eqref{5r}$ and
adiabatic index $\eqref{54y}$ with radius $r$.}
\end{figure}

\subsubsection{Adiabatic Index}

This index is key to determining the dynamical stability of
relativistic stars under infinitesimal adiabatic perturbations, a
concept firstly explored by Chandrasekhar \cite{50y}. For a
relativistic system to be stable, the adiabatic index must exceed
4/3 \cite{51y}. If it approaches or drops below this threshold, the
star can become dynamically unstable, potentially leading to
gravitational collapse \cite{52y}. Its expression is
\begin{equation}\label{54y}
\Delta =\bigg(1+\frac{\rho}{P}\bigg)\frac{dP}{d\rho},
\end{equation}
which is the ratio of pressure change to the change in density
during an adiabatic process. In the interior region where \(P =
-\rho\), \(\Delta = 0\), indicating no significant change in
pressure relative to density. In the intermediate region where \(P =
\rho\), \(\Delta = 2\), indicating a typical relationship where
pressure increases with density. The graphical behavior is shown in
Figure \textbf{9}. As the radius \(\mathbb{R}\) increases, the
growing pressure-density change suggests greater stability as the
gravastar expands.

\section{Final Remarks}

This paper presents exact solutions for an isotropic fluid gravastar
within a spherically symmetric static spacetime, building on the
work of Mazur and Mottola \cite{3a} in Rastall theory. Gravastars,
which resemble black holes, address many of their associated issues.
Our focus is on the effects of EMT's non-conservation by applying
Rastall FEs. Using a static sphere with the Kuchowicz potential, we
model physically viable, singularity-free gravastar structure. These
solutions are free of singularities and event horizons, with
Rastall's parameter affecting the interior, shell, and exterior
structure of the gravastar. The balance between outward pressure and
inward gravitational force prevents singularity formation at the
core. Our analysis reveals that the gravitational mass takes a
positive value in the inner domain, but at the core, it is zero.
Figure \textbf{1} shows the regular, positive profile of the mass
function in the inner region. The metric functions within the
interior and across the thin shell remain singularity-free. Figure
\textbf{2} shows the variation of \( e^{\psi(r)} \), confirming the
solution's physical regularity and acceptability. The physical
features, including shell thickness dependence, are discussed in
detail, with graphical representations. Some key findings are
summarized as follows.
\begin{itemize}
\item \textbf{Pressure-density relation:} Within the gravastar's interior
region, negative pressure remains consistently negative and both
pressure and energy density maintain constant values. Figure
\textbf{3} specifically illustrates how the pressure of the
ultra-relativistic fluid within the shell varies as a function of
the radial coordinate $r$.

\item \textbf{Junction interface:} The junction condition for thin shell
formation between inner and outer spacetimes is analyzed. Following
Israel's conditions, we investigate the variation of surface energy
density and surface pressure, as shown in Figure \textbf{4} for
different Rastall parameter values. Both $\Xi$ and $\Upsilon$
exhibit positive behavior, with the gravastar mass
$M=3.75M_{\odot}$, $\mathbb{R}_1=10.00km$ and
$\mathbb{R}_2=10.10km$. This illustrates how the gravastar boundary
varies with the Rastall parameter.

\item \textbf{EoS parameter:} The EoS parameter in the shell region
reveals that, for large radius (\(\mathbb{R}\)), the EoS approaches
the dark energy regime as the Rastall parameter \(\varkappa \to 0\).
For small radius, however, \(\mathcal{W}(\mathbb{R}) \to
-\varkappa\) and thus, for \(\varkappa \to 0\),
\(\mathcal{W}(\mathbb{R}) \to 0\), which corresponds to a dust shell
configuration. In Figure \textbf{5}, for each value of the parameter
\(\varkappa\), the EoS value at a given \( \mathbb{R} \) is slightly
different, but the overall increasing behavior is maintained.

\item \textbf{Proper length of the shell:} Figure \textbf{6} depicts how the
length ($\mathcal{L}$) of thin shell in a gravastar structure varies radially.
The plot reveals a linear increase in length as the shell thickness grows.

\item \textbf{Energy of the shell:} Figure \textbf{7} reveals that the energy rises
within the shell exhibits a positive and monotonically increasing trend
as it approaches the outer surface. This indicates that the outer boundary
of the shell is denser than its inner edge.

\item \textbf{Entropy of the shell:} Figure \textbf{8} illustrates the radial
variation of the shell entropy, showing that the increase in entropy with the
shell's thickness is a result of the system's thermodynamic behavior.
However, this behavior does not indicate any instability in our solution.

\item \textbf{Gravitational surface redshift:} The model's stability is further validated
by examining the gravitational surface redshift, which adheres to
the Buchdahl limit for isotropic stellar configurations \cite{83z}.
From Figure \textbf{9}, it can be deduced that our gravastar
structure is both stable and physically valid.

\item \textbf{Adiabatic index:} In Figure \textbf{9}, the increasing trend of the
adiabatic index suggests that the gravastar shell remains stable,
while also pointing to a potential instability in the inner region
of the gravastar.
\end{itemize}

Several astrophysical tests could detect the existence of gravastar
models. We discuss these tests and how they get affected under the
considered non-conserved theory in the following.
\begin{itemize}
\item Gravastar shadows are similar to black hole shadows, appearing
as dark regions against brighter emissions but without an event
horizon due to the gravastar's compact nature. Under Rastall
gravity, these shadows could exhibit variations in size and shape.
Such differences are essential for distinguishing gravastars from
black holes and for evaluating the accuracy of Rastall theory's
predictions.
\item Microlensing involves a massive object, such as a gravastar,
amplifying light from a distant source as it passes between the
source and an observer, revealing the presence of compact masses
indirectly. Under the considered gravity, changes in gravitational
interactions could alter the intensity and pattern of microlensing
light curves. These deviations from expected patterns could provide
unique signatures that distinguish gravastars from black holes.
\item The EHT is a network of global radio
telescopes that takes high-resolution images of the regions around
supermassive black hole candidates, capturing the shadow and light
rings formed by intense gravitational bending. If gravastars are
real and influenced by Rastall theory, the EHT might detect unusual
ring-like structures around these objects, differing from typical
black hole observations.
\item LIGO detectors detect spacetime ripples from
massive astrophysical events, such as mergers or collapses. If
gravastars merge, they too could produce detectable gravitational
waves. These waves might display unique characteristics, like
altered waveforms or energy distributions, under Rastall gravity,
potentially distinguishing them from black hole signals.
\end{itemize}
In conclusion, this study offers a novel approach to modeling
gravastars by employing the Kuchowicz metric potentials within a
spherical spacetime. Although the use of Rastall parametric values
helps to determine the physically admissible results for the
gravitational vacuum structure, our analysis confirms that the
resulting solutions are regular, finite, and well-behaved at the
origin. Consequently, the proposed gravastar model is both
theoretically consistent and physically viable. These results
encourage further investigations using other modified gravitational
frameworks to expand the theoretical landscape of gravastar models.
\\\\
\textbf{Data Availability Statement:} The research presented in this
paper did not utilize any data.

\end{document}